\documentclass[aps,prb,twocolumn,10pt,superscriptaddress,notitlepage]{revtex4-1}

\newcommand{\papertitle}{Near-field photocurrent nanoscopy on bare and encapsulated graphene}

\usepackage[utf8]{inputenc}
\usepackage{graphicx}

\usepackage{graphicx}
\usepackage{amsmath}
\usepackage{natbib}
\usepackage{float}
\usepackage{siunitx}
\usepackage{lmodern}
\usepackage[T1]{fontenc}
\usepackage[unicode=true,
	    colorlinks=true,
	    linkcolor=blue,
	    citecolor=blue,
	    urlcolor=blue]{hyperref} 
\usepackage{sansmath,caption}
\newcommand{\newtextbar}[1][3]{\scalebox{#1}[1]{\textbar}}
\DeclareCaptionLabelSeparator{vertline}{\textbf{~\newtextbar}}
\usepackage[font=sf, labelfont={sf,bf},size=small,justification=RaggedRight,textfont={sf,sansmath},labelsep=vertline]{caption}
\usepackage{ulem}

\newcommand{\Ipc}{\ensuremath{I_\mathrm{PC}}}
\newcommand{\Vbg}{\ensuremath{V_\mathrm{BG}}}
\newcommand{\lcool}{\ensuremath{l_\mathrm{cool}}}
\newcommand{\SGB}{\ensuremath{S_\mathrm{GB}}}
\newcommand{\SG}{\ensuremath{S_\mathrm{G}}}

\newcommand{\sopt}{\ensuremath{\xi_{\mathrm{opt}}}}   

\begin{document}
\title{\Large\textsf{\papertitle}}


\author{Achim Woessner}
\affiliation{\footnotesize ICFO – Institut de Ciències Fotòniques, Mediterranean Technology Park, 08860 Castelldefels (Barcelona), Spain}
\author{Pablo Alonso-González}
\thanks{These authors contributed equally}
\affiliation{\footnotesize CIC nanoGUNE, 20018 Donostia-San Sebastian, Spain}
\author{Mark B. Lundeberg}
\thanks{These authors contributed equally}
\affiliation{\footnotesize ICFO – Institut de Ciències Fotòniques, Mediterranean Technology Park, 08860 Castelldefels (Barcelona), Spain}
\author{Yuanda Gao}
\affiliation{\footnotesize Department of Mechanical Engineering, Columbia University, New York, NY 10027, USA}
\author{Jose E. Barrios-Vargas}
\affiliation{\footnotesize ICN2 – Institut Català de Nanociència i Nanotecnologia, Campus UAB, 08193 Bellaterra, Spain}
\author{Gabriele Navickaite}
\affiliation{\footnotesize ICFO – Institut de Ciències Fotòniques, Mediterranean Technology Park, 08860 Castelldefels (Barcelona), Spain}
\author{Qiong Ma}
\affiliation{\footnotesize Department of Physics, Massachusetts Institute of Technology, Cambridge, MA 02139, USA}
\author{Davide Janner}
\affiliation{\footnotesize ICFO – Institut de Ciències Fotòniques, Mediterranean Technology Park, 08860 Castelldefels (Barcelona), Spain}
\author{Kenji Watanabe}
\affiliation{\footnotesize National Institute for Materials Science, 1-1 Namiki, Tsukuba 305-0044, Japan}
\author{Aron W. Cummings}
\affiliation{\footnotesize ICN2 – Institut Català de Nanociència i Nanotecnologia, Campus UAB, 08193 Bellaterra, Spain}
\author{Takashi Taniguchi}
\affiliation{\footnotesize National Institute for Materials Science, 1-1 Namiki, Tsukuba 305-0044, Japan}
\author{Valerio Pruneri}
\affiliation{\footnotesize ICFO – Institut de Ciències Fotòniques, Mediterranean Technology Park, 08860 Castelldefels (Barcelona), Spain}
\affiliation{\footnotesize ICREA – Institució Catalana de Recerca i Estudis Avançats, 08015 Barcelona, Spain}
\author{Stephan Roche}
\affiliation{\footnotesize ICN2 – Institut Català de Nanociència i Nanotecnologia, Campus UAB, 08193 Bellaterra, Spain}
\affiliation{\footnotesize ICREA – Institució Catalana de Recerca i Estudis Avançats, 08015 Barcelona, Spain}
\author{Pablo Jarillo-Herrero}
\affiliation{\footnotesize Department of Physics, Massachusetts Institute of Technology, Cambridge, MA 02139, USA}
\author{James Hone}
\affiliation{\footnotesize Department of Mechanical Engineering, Columbia University, New York, NY 10027, USA}
\author{Rainer Hillenbrand}
\affiliation{\footnotesize CIC nanoGUNE and UPV/EHU, 20018 Donostia-San Sebastian, Spain}
\affiliation{\footnotesize IKERBASQUE, Basque Foundation for Science, 48011 Bilbao, Spain}
\author{Frank H.L. Koppens}
\email{frank.koppens@icfo.eu}
\affiliation{\footnotesize ICFO – Institut de Ciències Fotòniques, Mediterranean Technology Park, 08860 Castelldefels (Barcelona), Spain}

\begin{abstract}\textsf{\textbf{\begin{normalsize}\begin{center}\begin{minipage}{\textwidth}
Opto-electronic devices utilizing graphene have already demonstrated unique capabilities, which are much more difficult to realize with conventional technologies. 
However, the requirements in terms of material quality and uniformity are very demanding.
A major roadblock towards high-performance devices are the nanoscale variations of graphene properties, which strongly impact the macroscopic device behaviour.  
Here, we present and apply opto-electronic nanoscopy to measure locally both the optical and electronic properties of graphene devices.
This is achieved by combining scanning near-field infrared nanoscopy with electrical device read-out, allowing infrared photocurrent mapping at length scales of tens of nanometers. 
We apply this technique to study the impact of edges and grain boundaries on spatial carrier density profiles and local thermoelectric properties.
Moreover, we show that the technique can also be applied to encapsulated graphene/hexagonal boron nitride (h-BN) devices, where we observe strong charge build-up near the edges, and also address a device solution to this problem.
The technique enables nanoscale characterization for a broad range of common graphene devices without the need of special device architectures or invasive graphene treatment. 
\end{minipage}\end{center}\end{normalsize}}}\end{abstract}

\maketitle


\begin{figure*}[t]
\centering
\includegraphics{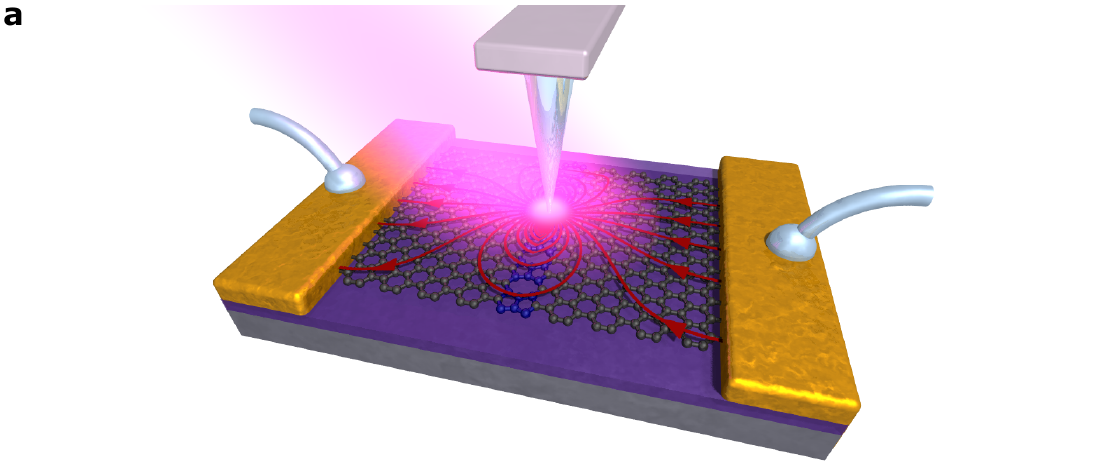} %
\includegraphics{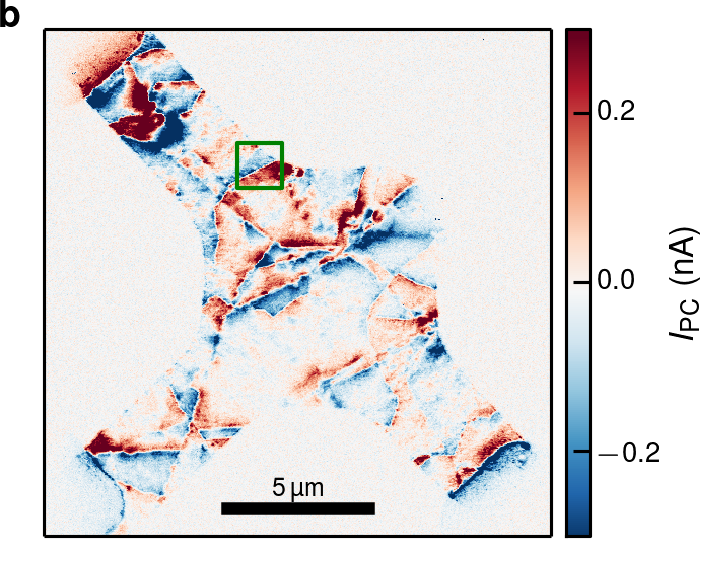}
\includegraphics{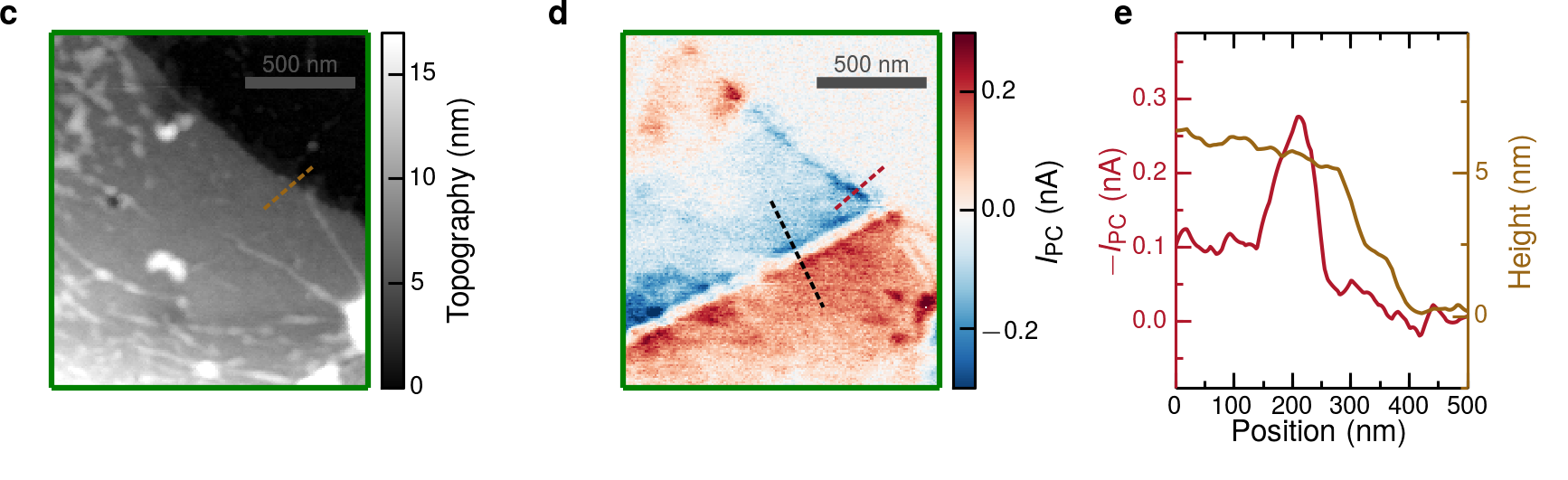}
\caption{
\label{fig1}
\textbf{Near-field photocurrent working principle and photocurrent from grain boundaries.}
\textbf{a},
Sketch of the scattering-type scanning near-field optical microscope setup. 
A mid-infrared laser illuminates the atomic force microscope tip, which generates a locally concentrated optical field, which is absorbed by the graphene generating a position dependent photocurrent.
The blue region in the graphene lattice represents a grain boundary with a modified Seebeck coefficient.
The arrows sketch the photocurrent flow pattern.
For each position only the magnitude and direction of the current are measured.
The sketch is not to scale.
\textbf{b},
$\Ipc$ map at at backgate voltage $\Vbg=0$~V of a CVD graphene device with three contacts: top left (drain), right (source) and bottom left (ground).
Both grain boundaries and wrinkles show characteristic photocurrent patterns.
The green box indicates the measurement region in \textbf{c},\textbf{d}.
\textbf{c},
Topography of etched CVD graphene does not show grain boundary but only wrinkles and other inhomogeneities due to the transfer process.
\textbf{d},
$\Ipc$ at $\Vbg=0$~V clearly shows a grain boundary and the expected sign change around it.
The black dashed line indicates the measurement positions in Fig.~\ref{fig2}a,d.
\textbf{e},
Topography (brown) and $\Ipc$ (red) measured at the brown line in \textbf{c} and the red line in \textbf{d} respectively. 
} \end{figure*}

As large scale integration and wafer scale device processing capabilities of graphene have become available,
\cite{Li2009a,Reina2009,Bae2010,Bonaccorso2012a,Ren2014,Lee2014c,Gao2014,DeHeer2011}
technological implementations of electronic and opto-electronic graphene devices are within reach.\cite{Ferrari2014,Akinwande2014,Koppens2014} 
At the same time, to achieve high device performance, any imperfections at the nanometer or even atomic scale need to be minimized or even eliminated.
For example, in large area graphene, grown by chemical vapour deposition (CVD), grain boundaries are the stitching regions between different mono-crystalline parts of graphene and act as carrier scatterers, limiting the graphene mobility and uniformity.\cite{Yazyev2010,Yu2011}
These nanoscale defects are elusive to many standard characterization techniques without special treatment of the graphene.\cite{Cummings2014,Yazyev2014}
In addition, even perfectly monocrystalline graphene is still highly sensitive to its environment, and on typical substrates charge-density inhomogeneities (charge puddles)\cite{Martin2007,Chen2008,Zhang2009a,Decker2011,Xue2011,Burson2013} and additional doping near contacts, defects and edges arise, which reduce the device performance as well.
Therefore it is important to efficiently probe the nanoscale opto-electronic properties of graphene and to understand the microscopic physical behaviour.

A major challenge is that many of the available techniques are invasive,\cite{Duong2012a} rely on specifically designed device structures,\cite{Yu2011,Huang2011b,Yasaei2015} image only very small areas,\cite{Martin2007,Chen2008,Deshpande2009,Gibertini2012,Huang2011b,Cho2013} rely on high doping of the graphene,\cite{Fei2013} need unhindered electrical access of the probe to the graphene,\cite{Martin2007,Chen2008,Deshpande2009,Gibertini2012,Huang2011b}  or lack the desired nanometer resolution\cite{Ferrari2013} and are difficult to implement.
Therefore, a nanoscopic tool that probes both electrical and optical response of graphene devices at nanometer length scales is highly desired.

Here we demonstrate fully non-invasive room-temperature scanning near-field photocurrent nanoscopy\cite{Mueller2009,Mauser2014,Mauser2014a,Grover2015} for the first time applied with infrared frequencies and use it to study the nanoscale opto-electronic properties of devices that can later be used for real applications. 
This technique allows measuring the properties of graphene devices that affect their performance with high spatial resolution in atmospheric conditions.
We apply this technique to study the microscopic physics of grain boundaries and charge density inhomogeneities.
In addition, we study encapsulated graphene devices,\cite{Wang2013,Kretinin2014} where the encapsulation would prevent many other scanning probe techniques from accessing local properties of graphene.
In general, this technique operates most effectively with mid-infrared light because it does not lead to photodoping\cite{Ju2014} and it is more stable in operation, compared to visible light. 

The measurement principle is sketched in Fig.~\ref{fig1}a.
The setup is based on a scattering-type scanning near-field optical microscope (s-SNOM)\cite{Fei2011,Fei2013} augmented with electrical contact to the sample to measure currents in situ.
In contrast to conventional s-SNOM we do not need to measure the outscattered light but rather directly measure current induced by the near-field as explained in the following. 
A \SI{10.6}{\micro\metre} mid-infrared laser illuminates a metallized atomic force microscope probe, tapping at its mechanical resonance frequency.
Part of the incoming light, polarized parallel to the shaft of the probe, excites a strong electric field at the tip apex due to an antenna effect.\cite{Keilmann2004a}
The spatial extent of this near-field is on the order of \SI{25}{\nano\metre}, limited only by the tip radius and much smaller than the free space wavelength of the impinging light.\cite{Keilmann2004a}

The near and far fields impinging on the device induce charge flows in the device (by mechanisms discussed below), and drive currents into an external current amplifier via contacts on the device.
We isolate the part of the current that is induced by near fields by demodulating the current at the second harmonic of the tip tapping frequency.\cite{Keilmann2004a} 
This demodulated current is denoted $\Ipc$ and referred to as near-field photocurrent and is obtained together with near-field optical and topography information.
A typical map of $\Ipc$, obtained by scanning the tip over a CVD graphene device, is shown in Fig.~\ref{fig1}b.

We can assess the spatial resolution of the photocurrent maps by comparing a region near the edge (Fig.~\ref{fig1}d) with a topographic image from the same region (Fig.~\ref{fig1}c). 
As can be seen, $\Ipc$ falls to zero for tip locations away from the graphene on a similar length scale as the topography, demonstrating the successful isolation of near field contributions.
In Fig.~\ref{fig1}e we quantify the resolution by observing the change in $\Ipc$ as the tip is moved over the edge of graphene.
The full width at half maximum of the photocurrent peak at this location is $\sim$~100~nm, matching the rise distance in the topographic signal.
This resolution is far below any limits relating to the \SI{10.6}{\micro\metre} free space light wavelength.

As to the physical mechanism of the photocurrent, we consider the photo-thermoelectric effect that has been shown to dominate the photoresponse of graphene:\cite{Lemme2011,Gabor2011,Song2011b,Tielrooij2014,Badioli2014,Koppens2014}
the light (in this case, the tip-enhanced near-field) locally heats the graphene, and this heat acts via non-uniformities in Seebeck coefficient $S$ to drive charge currents within the device and into the contacts (see Methods). 
Therefore, we interpret the variations of $\Ipc$ in terms of microscopic variations in $S$.
The Seebeck coefficient, which depends on material properties such as carrier density and mobility, is a measure of the electromotive force driven by a temperature difference in a material.
A complete description of $\Ipc$ needs to take into account the carrier cooling length\cite{Gabor2011,Song2011b} and overall sample geometry.\cite{Song2014}
The carrier cooling length $\lcool = \sqrt{\kappa/g}$, where $\kappa$ the thermal conductivity in plane and $g$ the thermal conductivity out of plane to the heat sinking substrate, describes how far heat propagates through the charge carriers, before dissipating to the environment.\cite{Song2011b}
A quantitative model of the thermoelectric photocurrent mechanism can be found in the Methods section and in the Supplement.

\begin{figure}[t]
\includegraphics{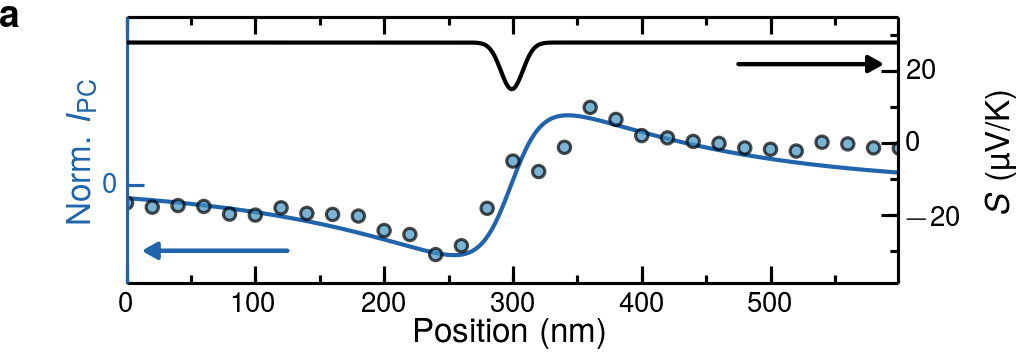}
\includegraphics{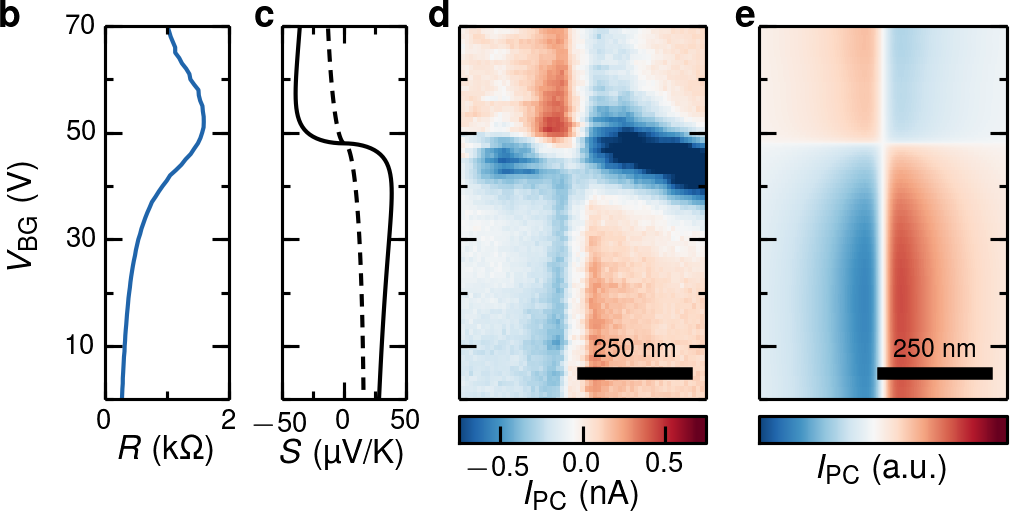}
\caption{
\label{fig2}
\textbf{Photocurrent profile at a grain boundary and its gate voltage dependence.}
\textbf{a}, Photocurrent profile, measured at the black dashed line in Fig.~\ref{fig1}d, perpendicular to the grain boundary at $\Vbg=0$~V shows good agreement with the photo-thermoelectric model with $\lcool = 140$~nm.
The modelled spatial Seebeck profile (with FWHM 20~nm) is shown in black.
\textbf{b}, Two-probe device resistance as a function of $\Vbg$.
\textbf{c}, Simulated Seebeck coefficient $S_\mathrm{G}$ for pristine graphene (solid line) and $S_\mathrm{GB}$ for polycrystalline graphene with an average grain size of 25~nm (dashed line) (see Supplement, II~D).
\textbf{d}, Backgate dependent photocurrent profile perpendicular to the grain boundary shows that the grain boundary changes its sign at the charge neutrality point.
\textbf{e}, Simulated backgate dependent photocurrent profile based on the Seebeck profiles in \textbf{c}.
}\end{figure}

\begin{figure}[b]
\centering
\includegraphics{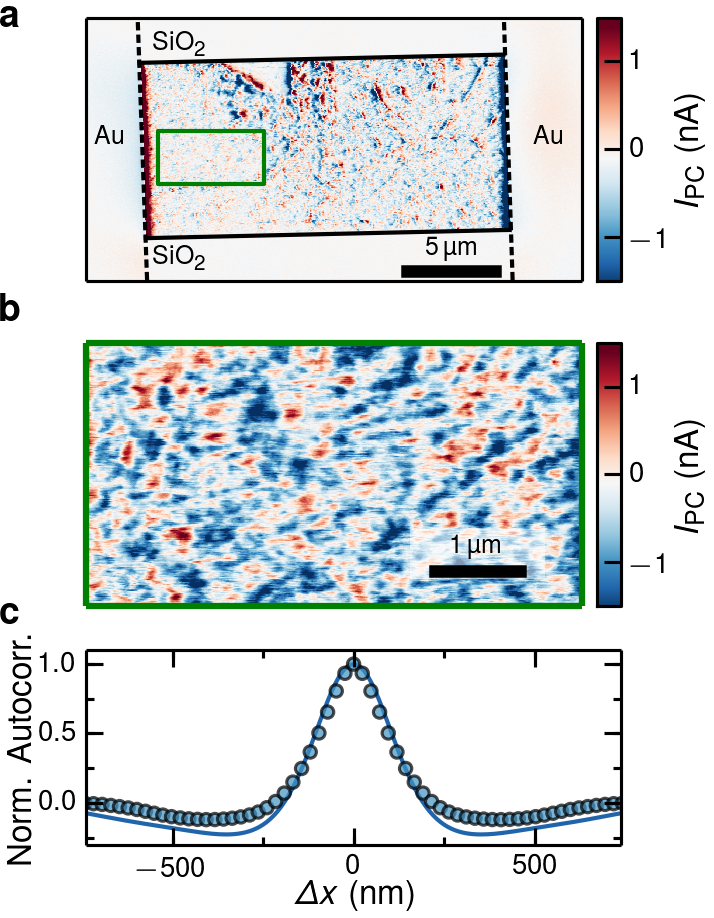}%
\caption{
\label{fig3}
\textbf{Photocurrent from charge puddles.}
\textbf{a}, Near-field photocurrent map of an exfoliated graphene device on 300~nm SiO$_2$ at $\Vbg=20$~V.
The dashed lines indicate the position of the contacts and solid lines the graphene edges.
The green box indicates the measurement region in \textbf{b}.
\textbf{b}, Detailed photocurrent map at the charge neutrality point of the device ($\Vbg=7$~V) reveals the charge puddles and the high spatial resolution of the technique.  
\textbf{c}, Autocorrelation of the photocurrent from charge puddles at $V_\mathrm{D}$ (data points) compared to photocurrent expected from a random charge puddle distribution and $\lcool = 200$~nm (blue curve).
Autocorrelation is taken along the source drain current path.
}\end{figure}

\begin{figure*}[t]
\centering
\includegraphics{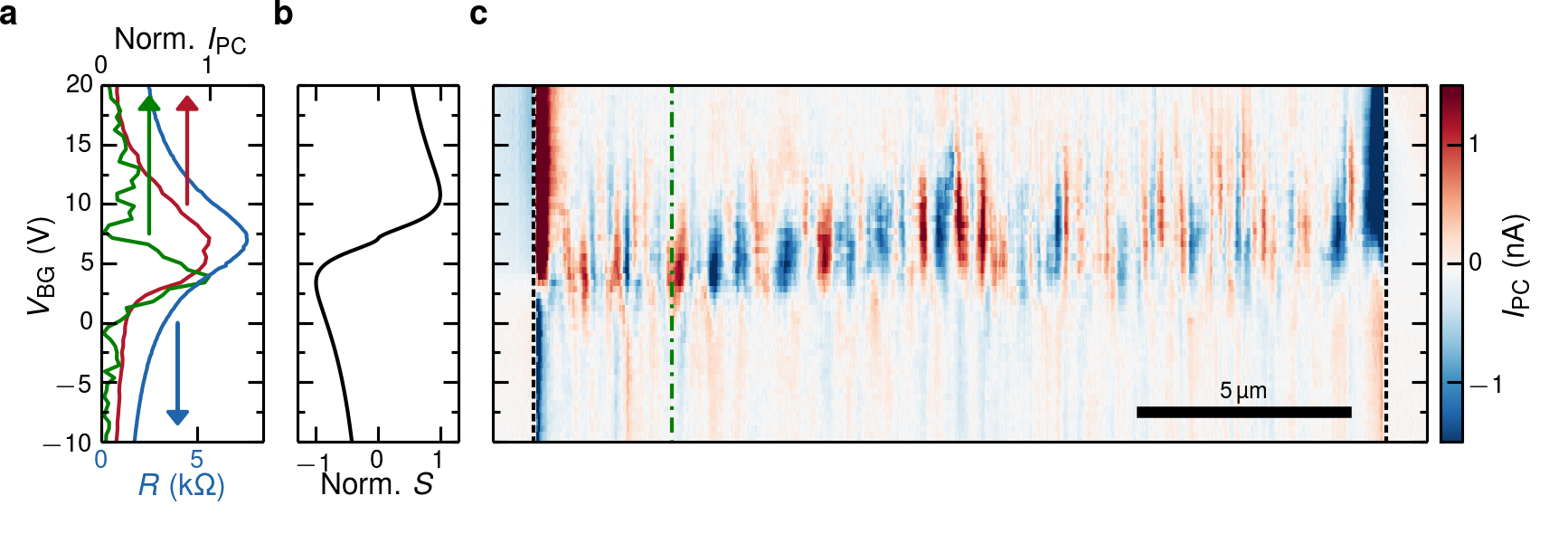}%
\caption{
\label{fig4}
\textbf{Dependence of photocurrent profiles on backgate voltage reveals doping inhomogeneities.}
\textbf{a}, Backgate dependence of the resistance of the device measured simultaneously to the photocurrent in blue.
The red curve shows the normalized root mean square of the photocurrent across the device.
The green curve shows a single normalized photocurrent backgate trace, corresponding to the green dashed dotted line in \textbf{c}. 
\textbf{b},
Backgate dependent Seebeck coefficient of graphene, calculated from the gate dependent resistance in \textbf{a} using the Mott formula.\cite{Zuev2009a}
\textbf{c}, Backgate dependence of the photocurrent across the device. 
Graphene is between the black dashed lines, which indicate the edges of the metal contacts.
}\end{figure*}

We first discuss the application of this infrared near-field photocurrent technique to grain boundaries, which are responsible for some of the line-shaped features in the photocurrent map in Fig.~\ref{fig1}b.
The region within the green frame is shown with higher resolution in Fig.~\ref{fig1}d, exhibiting a strong photocurrent signal that changes sign along a sharp boundary, yet the graphene is topographically flat in the vicinity of this boundary (Fig.~\ref{fig1}c).
We show now that this type of feature indicates a grain boundary.

Figure \ref{fig2}a shows a line profile of $\Ipc$ across the boundary feature identified in Fig.~\ref{fig1}d.
This antisymmetric $\Ipc$ can be explained by a localized deviation in $S$ at the boundary, i.e. a line defect within an otherwise uniform thermoelectric medium.
Indeed, grain boundaries behave as localized lines of strongly modified electronic properties, within otherwise uniform graphene.\cite{Huang2011b,Duong2012a,Tsen2012,Fei2013,VanTuan2013}
We remark that the decay of the photocurrent away from the boundary extends over more than 100~nm, which is due to a larger hot carrier cooling. 
We find in this case $\lcool=140$~nm.

To gain more insight in the Seebeck coefficient at the grain boundary, we tune the carrier density by a global gate (Fig.~\ref{fig2}d).
We observe that the antisymmetric spatial photocurrent profile changes sign as the backgate voltage $\Vbg$ passes the peak in resistance, i.e., the global charge neutrality point $V_\mathrm{D}$.
The Seebeck coefficient $S_\mathrm{G}$ of graphene itself changes sign at the charge neutrality point\cite{Zuev2009a,Wei2009a,Gabor2011,Lemme2011} (Fig.~\ref{fig2}c), thus our data implies that the Seebeck coefficient of the grain boundary $S_\mathrm{GB}$ is always smaller in magnitude than $S_\mathrm{G}$, since $\Ipc(\Vbg) \propto S_\mathrm{G}(\Vbg) - S_\mathrm{GB}(\Vbg)$.

Using a polycrystalline graphene model, we compute the resistance due to grain boundaries using a Kubo transport formalism and real space simulations.\cite{Cummings2014} 
$S_\mathrm{GB}$ is the ratio of the first- and zero-order Onsager coefficients (see Supplement, II~D).
Indeed we find that $S_\mathrm{GB}$ is always smaller in magnitude and has a similar lineshape as $S_\mathrm{G}$ in the carrier density range measured (Fig.~\ref{fig2}c).
Fig.~\ref{fig2}e shows a simulation of the photocurrent for the calculated Seebeck coefficients, which is in excellent agreement with the measurements.

We next examine near-field photocurrent in a typical two probe exfoliated graphene device (Fig.~\ref{fig3}).
A strong photocurrent is obtained with the tip near the metal contacts, similar to previous near- and far-field measurements.\cite{Lee2008,Mueller2009,Tielrooij2014}
Additionally, an apparently random pattern of photocurrent is present throughout the device, as in high-resolution far-field measurements\cite{Lee2008} but at a much finer scale.

The random photocurrent pattern between the contacts in Fig.~\ref{fig3}a indicates random variations in Seebeck coefficient over short length scales.
Random variations of the Seebeck coefficient are indeed expected since it depends on carrier density,\cite{Zuev2009a} which in turn has fine-scaled inhomogeneities (charge puddles).\cite{Martin2007,Chen2008,Zhang2009a,Decker2011,Xue2011}
The photocurrent variations can thus be used to gain insight in the charge puddle distribution.
A more detailed view of the photocurrent due to charge puddles in Fig.~\ref{fig3}b shows that the length scale that can be resolved is on the order of hundreds of nanometers.

Quantitatively, from the autocorrelation of the photocurrent in comparison to a photo-thermoelectric model taking into account the size of the charge puddles in Fig.~\ref{fig3}c we extract $\lcool \sim$~\SI{200}{\nano\metre}.
The charge puddles are modelled to have a size of $\sim$ 20~nm, in accordance with measurements of graphene on SiO$_2$.\cite{Chen2008,Zhang2009a,Decker2011,Xue2011}

By changing the gate voltage we study the carrier density profile with high spatial resolution (see Fig.~\ref{fig4}) and highlight the possibility of spatially resolving the charge neutrality point for a large device.
$\Ipc$ from charge puddles is largest around the charge neutrality point and varies with position.
This is consistent with the very high sensitivity of the Seebeck coefficient to changes in carrier density, near zero density (Fig.~\ref{fig4}b).
This allows us to map the local carrier density offset (charge inhomogeneity) throughout the device, as indicated by the extremum of photocurrent in a scan of photocurrent vs. gate voltage (Fig.~\ref{fig4}c).
In contrast to the grain boundary photocurrent we do not observe the charge puddle photocurrent change sign when sweeping the average carrier density through the charge neutrality point, as expected from the gate dependence the of Seebeck coefficient.

We can thus resolve the local charge neutrality point at a given position of the device (green curve, Fig.~\ref{fig4}a), which can be different from the global charge neutrality point $V_\mathrm{D}$, the backgate voltage $\Vbg$ at which the resistance is maximum (blue curve, Fig.~\ref{fig4}a).
We show that the global charge neutrality point (blue curve, Fig.~\ref{fig4}a) is determined by an average of the gate voltages at which the local charge neutrality points appear (red curve, Fig.~\ref{fig4}a).
Spatially resolved puddle photocurrent can be much narrower (green curve, Fig.~\ref{fig4}a) than the average of all possible current paths (red curve, Fig.~\ref{fig4}a)
This indicates that the graphene locally has less inhomogeneity. 
Thus the technique gives insight not only in the global but also in the local behaviour of the device.


\begin{figure*}[t]
\centering
\includegraphics{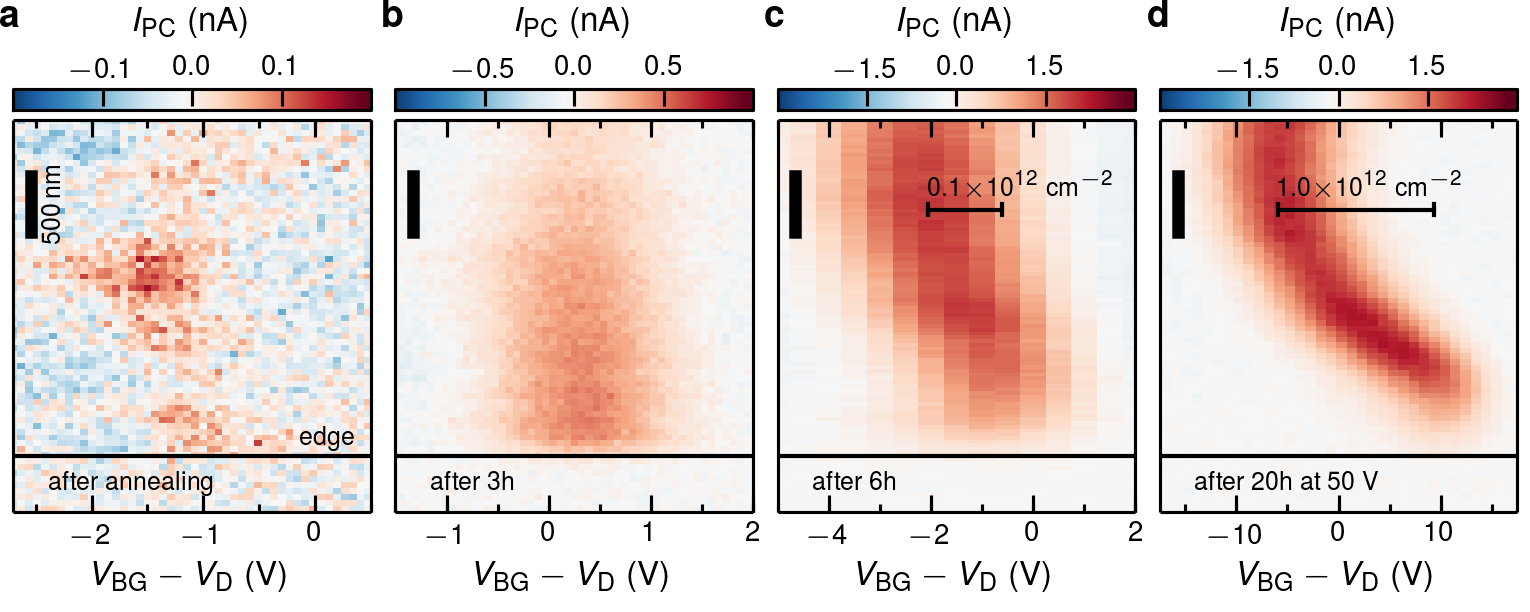}
\includegraphics{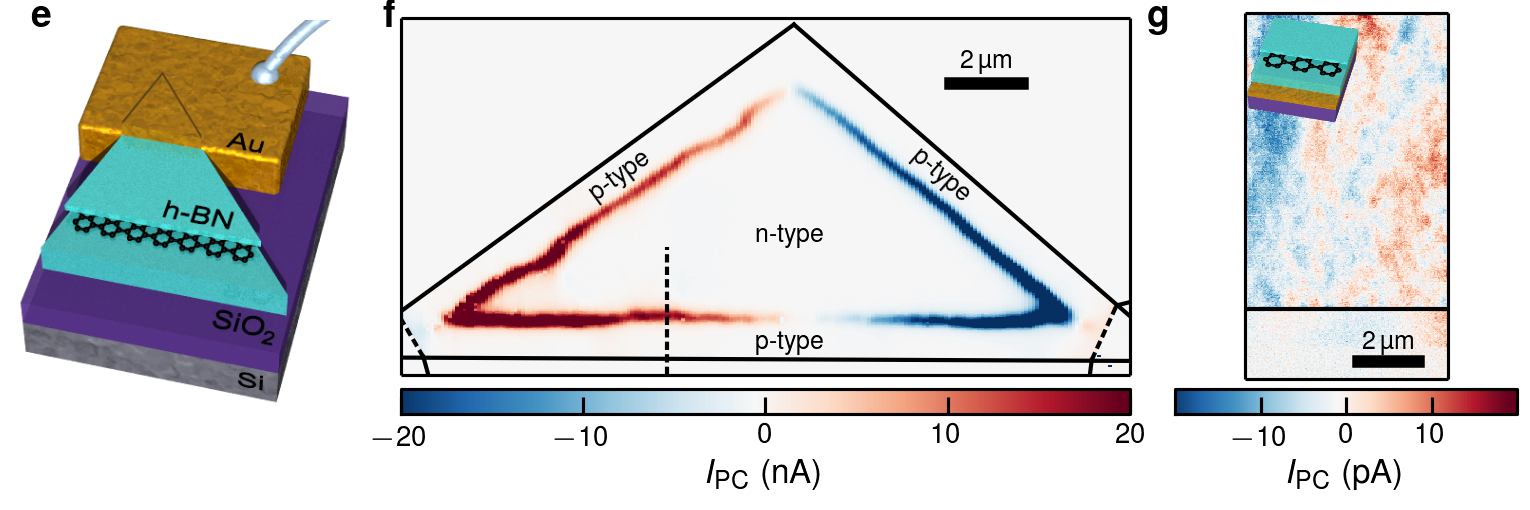}
\caption{
\label{fig5}
\textbf{Near-field photocurrent maps revealing edge doping in encapsulated graphene.}
\textbf{a},
Spatial photocurrent profile vs. backgate voltage $\Vbg$ (minus voltage of the resistance maximum $V_\mathrm{D}$) near the edge of encapsulated graphene. 
These data are taken directly after annealing the device. 
\textbf{b}, The same scan on the same device after three hours in air, and \textbf{c}, after annealing and applying $\Vbg$ up to 3~V. 
\textbf{d}, The same scan after $\sim$~20~hours in air and after applying $\Vbg$ up to 50~V.
\textbf{e},
Sketch of the device, a stack of h-BN(46~nm)/Gr/h-BN(7~nm) on a Si/SiO$_\mathrm{2}$(300~nm) wafer used as global backgate.
\textbf{f},
Photocurrent close to the resistance maximum at $\Vbg=-28$~V shows a triangular photocurrent pattern, due to edge $p$-doping.
The dashed lines in \textbf{f} indicate where the stack is underneath the gold.
\textbf{g},
Photocurrent from charge puddles in encapsulated graphene on a metal gate close to the charge neutrality point (sketch in inset).
The layers are AuPd(15~nm)/h-BN(42~nm)/Gr/h-BN(13~nm).
The electrical contacts are on the left and right outside of this figure.
In \textbf{a}-\textbf{d},\textbf{f},\textbf{g} the black solid lines indicate the graphene edge.
}\end{figure*}

Finally we apply this technique to a graphene device encapsulated between two layers of h-BN, using the polymer-free van der Waals assembly technique\cite{Wang2013,Kretinin2014} as sketched in Fig.~\ref{fig5}e.
This device lies on top of an oxidized silicon wafer, used as a backgate.
The stack is etched into a triangle and electrically side-contacted by metal electrodes.\cite{Wang2013}  
Recent studies\cite{Shalom2015,Allen2015} have shown that the edges affect where current flows in the device, in particular near charge neutrality.
In the following we study the build up of edge doping and provide a solution to this.

While monitoring the photocurrent of such encapsulated devices, we observe indications of strong carrier density variations near the edges over micrometer scales.
These variations are influenced by lighting conditions, gate voltages, and temperature, and evolve over timescales ranging from minutes to weeks.
As an example, Figure~\ref{fig5}a-d shows a progression of photocurrent maps, taken after annealing the device at 200~$^{\circ}$C for 30 minutes to temporarily remove charge density variations near the edges.
Initially in Fig.~\ref{fig5}a we see very small photocurrents indicating a flat carrier density landscape.
After some time ($\sim$ hours), in the dark with only gate voltages smaller than 3~V applied, a small doping gradient between the contacts builds up.
This gradient leads to the stronger photocurrent shown in Fig.~\ref{fig5}b.
The local charge neutrality point, indicated by the maximum of photocurrent, is at the same position close to the edge of the device as further inside the bulk.
After keeping the device for 3 hours in ambient conditions we can see a change of the local charge neutrality point at the edge of the graphene compared to the bulk in Fig.~\ref{fig5}c.
The edge is slightly more $p$-type compared to the bulk.
Finally we apply high gate voltages, of in this case 50~V for $\sim$ 20 hours, to increase the edge doping.
A strong $p$-doping at the edge and an $n$-doping in the bulk of the graphene is induced in Fig.~\ref{fig5}d.
This indicates that electric field accelerates the speed and increases this type of edge doping.

We exploit the observed edge doping to create a natural $pn$-junction along the edge of the device.
For this we apply a backgate voltage at which the edge of the graphene is $p$-type and the bulk $n$-type.
We observe photocurrent at the junction in Fig.~\ref{fig5}f around the whole device, indicating that the edge doping is uniform around the graphene.
The photocurrent decays gradually towards the midline between the electrodes as a result of how the triangular geometry modifies the ability of the contacts to capture photocurrents.\cite{Song2014}
We are able to temporarily reset the edge doping by annealing the device on a hotplate at 200~$^{\circ}$C for 30 minutes.

While we have not been able to precisely identify the origin of the edge doping, we present here a technique to completely eliminate it.
We place encapsulated graphene on top of a local conductive gate, such as a 15~nm AuPd alloy in the case of Fig.~\ref{fig5}g.
We find that edge doping is efficiently suppressed even after extended periods of time at ambient conditions and high gate voltages.
Furthermore, such devices lack the photodoping effect observed for devices where the h-BN is in contact with SiO$_2$\cite{Ju2014} (see Supplement, I~D). 
We suspect that humidity that is able to penetrate between the boron nitride and the silicon oxide leading to trapped charges is responsible for the observed edge doping.

In the device with a metal gate we find small features due to charge puddles on top of a slowly varying background photocurrent, due to large scale carrier density inhomogeneities.
The size of the features due to charge puddles determined by autocorrelation is $\approx$ 800~nm. 
The long length scale of those features is either due to the longer cooling length of the encapsulated graphene compared to the graphene on SiO$_2$ or due to larger charge puddle size in the encapsulated devices.
Further work is required to clearly distinguish these effects.

To conclude, we have demonstrated that scanning near-field photocurrent nanoscopy is a versatile technique to characterize the electronic and opto-electronic and even previously inaccessible properties of relevant graphene devices.
This technique is highly promising for spatially resolved quality control of regular graphene devices without the need for special device structures and can therefore be readily applied.

\section*{\textsf{Methods}}
{\small
\subsubsection*{\textsf{Photocurrent model}}
Photocurrent $\Ipc$ in graphene as generated by the photo-thermoelectric effect and is described as:\cite{Song2011b,Gabor2011,Tielrooij2014}
\begin{equation*}
\Ipc = \frac{1}{RW} \iint \dfrac{\partial T}{\partial x} S \, dxdy
\end{equation*}
where $R$ is the total resistance including graphene, contacts and circuitry, $W$ the device width and $x$ the current flow direction.
This is valid for rectangular graphene devices and special care needs to be taken for arbitrary shapes, such as in Fig.~\ref{fig5}.\cite{Song2014}
For the temperature profile $T(x)$ we consider that the heat spreads in two dimensions with heat sinking to lattice and substrate, producing a $T(x)$ profile described by a modified Bessel function of the second kind, with a finite tip size correction (see Supplement, II~B).
A 25~nm finite tip size correction was used for all simulations.

\subsubsection*{\textsf{Measurement details}}
The s-SNOM used was a NeaSNOM from Neaspec GmbH, equipped with a CO$_2$ laser operated at \SI{10.6}{\micro\metre}, away from the phonon resonance of SiO$_2$, which can lead to strong substrate contributions to the photocurrent.\cite{Badioli2014}
The probes were commercially-available metallized atomic force microscopy probes with an apex radius of $\sim$~\SI{25}{\nano\metre}.
The tip height was modulated at a frequency of $\sim 250~\mathrm{kHz}$ with an amplitude of 60--80~nm.
A Femto DLPCA-200 current pre-amplifier was used.

\subsubsection*{\textsf{Device fabrication}}
The CVD graphene was transferred onto a self-assembled monolayer\cite{Chen2012a} on 285~nm of SiO$_2$ in order to stabilize the charge neutrality point.
The contacts were defined using optical lithography with Ti~(5~nm)/Pd~(35~nm).
The graphene was transferred onto deposited contacts.

The exfoliated graphene device was fabricated on a Si/SiO$_2$(300~nm) wafer, used as backgate. 
The Cr(0.8~nm)/Au(80~nm) contacts were defined using electron beam lithography.

The Si/SiO$_2$(300~nm)/h-BN(46~nm)/Gr/h-BN(7~nm) and the Si/SiO$_2$(300~nm)/AuPd(15~nm)/h-BN(42~nm)/Gr/h-BN(13~nm) stacks, were fabricated using the polymer-free van der Waals assembling technique.\cite{Wang2013} 
} 


\section*{\textsf{Acknowledgements}}
{\small
It is a great pleasure to thank Stijn Goossens, Klaas-Jan Tielrooij and Misha Fogler for many fruitful discussions and Fabien Vialla for the assistance in rendering the graphics in Fig.~\ref{fig1}a.
Open source software was used (www.matplotlib.org, www.python.org, www.povray.org).
F.H.L.K. acknowledges support by the Fundacio Cellex Barcelona, the ERC Career integration grant 294056 (GRANOP), the ERC starting grant 307806 (CarbonLight). 
F.H.L.K. and R.H. acknowledge support by the E.C. under Graphene Flagship (contract no. CNECT-ICT-604391).
D.J. acknowledges support from the Ramon y Cajal fellowship program.
Y.G. and J.H. acknowledge support from the US Office of Naval Research N00014-13-1-0662.
We acknowledge financial support from the Spanish Ministry of Economy and Competitiveness and ''Fondo Europeo de Desarrollo Regional'' through Grant TEC2013-46168-R.
Q.M. and P.J.H. have been supported by AFOSR Grant No. FA9550-11-1-0225 and the Packard Fellowship program. 
This work made use of the Materials Research Science and Engineering Center Shared Experimental Facilities supported by the National Science Foundation (NSF) (Grant No. DMR-0819762) and of Harvard's Center for Nanoscale Systems, supported by the NSF (Grant No. ECS-0335765).
S.R. acknowledges the Spanish Ministry of Economy and Competitiveness for funding (MAT2012-33911), the Secretaria de Universidades e Investigacion
del Departamento de Economia y Conocimiento de la Generalidad de Catalunya and the Severo Ochoa Program (MINECO SEV-2013-0295).
J.E.B.V. acknowledges support from SECITI (Mexico, D.F.).
}
%
\section*{\textsf{Author contributions}}
{\small
A.W. performed the experiments, analysed the data and wrote the manuscript.
P.A.-G. helped with measurements, interpretation and discussion of the results.
M.B.L. helped with data analysis, measurements, interpretation, discussion of the results and manuscript writing.
Y.G. fabricated the h-BN/graphene/h-BN devices.
G.N. and D.J. fabricated the CVD graphene devices.
Q.M. fabricated the exfoliated graphene devices.
K.W. and T.T. synthesized the h-BN.
J.E.B.V., A.W.C. and S.R. performed the simulations of the Seebeck coefficient at grain boundaries.
R.H. and F.H.L.K. supervised the work, discussed the results and co-wrote the manuscript.
All authors contributed to the scientific discussion and manuscript revisions.
}

\section*{\textsf{Competing Financial Interests}}
{\small
R.H. is co-founder of Neaspec GmbH, a company producing scattering-type scanning near-field optical microscope systems such as the ones used in this study. 
All other authors declare no competing financial interests.
}
\pagebreak
\newpage
\widetext
\cleardoublepage

\setcounter{equation}{0}
\setcounter{figure}{0}
\setcounter{table}{0}
\setcounter{page}{1}
\makeatletter
\renewcommand{\theequation}{s\arabic{equation}}
\renewcommand{\thefigure}{S\arabic{figure}}
\renewcommand{\thetable}{S\arabic{table}}
\renewcommand{\thepage}{\roman{page}}
\renewcommand{\bibnumfmt}[1]{[S#1]}
\renewcommand{\citenumfont}[1]{S#1}


\begin{center}
\textbf{\Large Supplementary Material: \papertitle}
\end{center}

\section{Measurements}
In this section we show measurements that further support the proof that the antisymmetric near-field photocurrent pattern of Fig.~1 of the main text indeed stems from a grain boundary.

\subsection{Topography, near-field optical scattering, and near-field photocurrent near a grain boundary}
\begin{figure*}[h]
\includegraphics{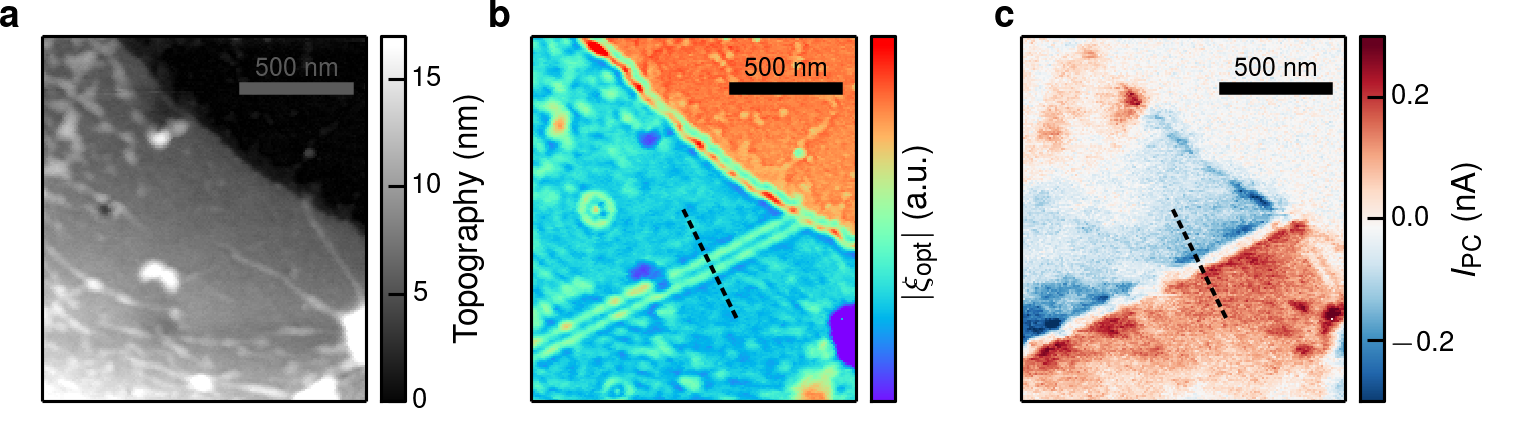}%
\caption{
\label{fig-supp-1}
\textbf{Comparison between topography, near-field optical scattering and near-field photocurrent.}
\textbf{a}, 
Topography of etched CVD graphene does not show grain boundaries but only wrinkles and other inhomogeneities due to the transfer process.
\textbf{b},
Near-field optical scattering shows the characteristic plasmonic double fringes around a grain boundary in CVD graphene at $\Vbg = 0$~V because the carrier density is $n_\mathrm{s} = 3.7\times 10^{12}$~cm$^{-2}$ and plasmons are supported.\cite{Fei2013,Schnell2014a}
\textbf{c}, 
Near-field photocurrent clearly shows the grain boundary and a sign change around it.
The dashed line in \textbf{b},\textbf{c} indicates where the backgate dependent measurements of the near-field optical scattering and near-field photocurrent in Figs.~2c and d respectively in the main text where taken.
All measurements of the near-field optical scattering $\sopt$ presented here were obtained from the third harmonic interferometric pseudo-heterodyne signal,\cite{Chen2012} measured with a cryogenic HgCdTe detector.
For simplicity Fig.~\ref{fig-supp-1}b only shows $|\sopt |$.
}\end{figure*}

In Fig.~1c and d of the main text, which are reprinted here as Fig.~\ref{fig-supp-1}a and c, we show the topography and the near-field photocurrent at a grain boundary for a backgate voltage of 0~V. 
To provide further evidence that the photocurrent stems from a grain boundary, we also show the simultaneously acquired near-field optical scattering data in Fig.\ref{fig-supp-1}b.
The near-field optical scattering data is acquired using a conventional scattering-type scanning near-field optical microscope (s-SNOM).\cite{Keilmann2004a}
It is a measure of the light that is scattered out of the graphene that interacts with the tip and is consequently detected in the far-field.
The near-field optical scattering data show the typical double fringe due to plasmon reflections at the grain boundary.\cite{Fei2013,Schnell2014a}

Even though conventional mid-infrared s-SNOM also offers the ability to detect grain boundaries in the near-field optical scattering,\cite{Fei2013} there are some distinct advantages to using the near-field photocurrent.
In order to observe plasmons due to grain boundaries, the graphene needs to be highly doped, as plasmons in graphene only propagate at at elevated carrier densities.\cite{Fei2012,Chen2012,Woessner2014}
For small carrier densities plasmons are heavily damped and no double fringe is visible, as observed by a decreasing visibility of the double fringe near-field optical scattering close to the charge neutrality point of the graphene (see Fig.~\ref{fig-supp-gb-gatedep}b).

In contrast, the near-field photocurrent shows a clear signature of the grain boundary even for small carrier densities (Fig.~\ref{fig-supp-gb-gatedep}c), which enables us to extract more information, as discussed in the main text. 
This demonstrates further that near-field photocurrent is a useful tool to provide insight into the local properties of graphene. 
The near-field photocurrent technique also facilitates measurements compared to measuring the near-field optical scattering as no weak outscattered light needs to be collected and no interferometric measurement is required.\cite{Ocelic2006,Schnell2014a}

When comparing the topography in Fig.~\ref{fig-supp-1}a with the near-field optical scattering in Fig.~\ref{fig-supp-1}b and the near-field photocurrent in Fig.~\ref{fig-supp-1}c it becomes evident that the near-field optical scattering and the near-field photocurrent contain information that is not visible in the topography.

\subsection{Gate dependence of a grain boundary}

\begin{figure*}[h]
\includegraphics{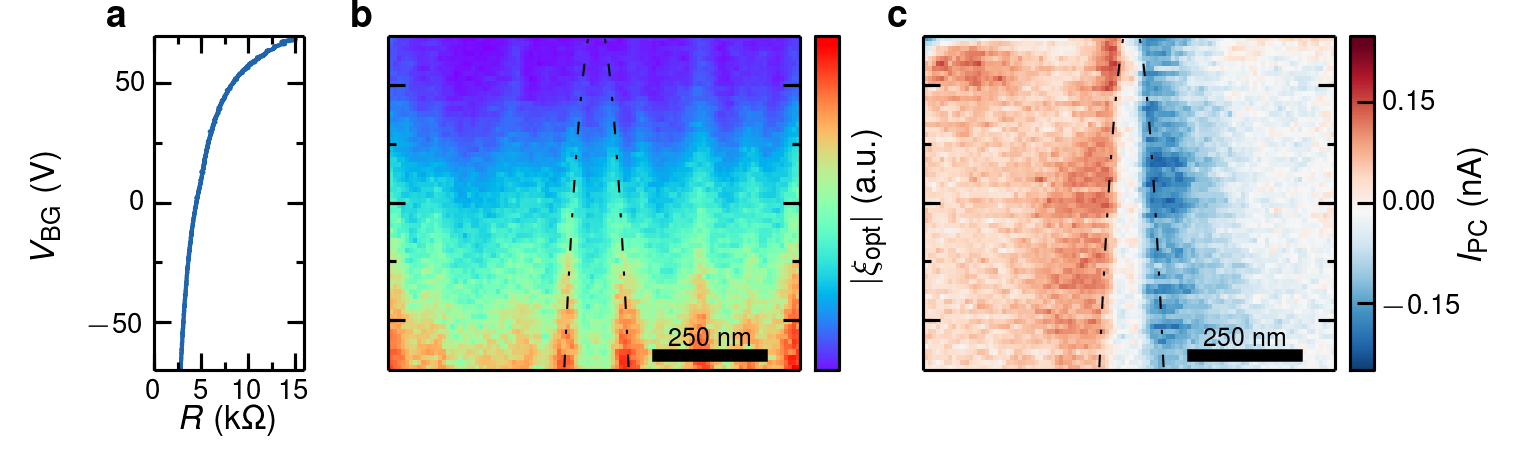}%
\caption{
\label{fig-supp-gb-gatedep}
\textbf{Gate dependent measurement of a grain boundary.}
\textbf{a}, 
Backgate dependence of the resistance of the device measured simultaneously to the near-field optical scattering and near-field photocurrent.
\textbf{b},
Near-field optical scattering shows the plasmon reflections due to the grain boundary and their dependence on the carrier density of the graphene.
\textbf{c},
Near-field photocurrent shows no sign change even for highly doped graphene.
The dashed dotted curves in \textbf{b} and \textbf{c} show the theoretical fringe spacing for a phase shift due to the reflection of $-3/4\pi$.\cite{Fei2013}
}\end{figure*}

As explained in the main text and shown in Fig.~2c, the Seebeck coefficient at or very near the grain boundary is smaller in magnitude than the Seebeck coefficient of the surrounding pristine graphene for all the carrier densities measured.
To further support this statement, we show the near-field optical scattering and near-field photocurrent for an extended carrier density range in Fig.~\ref{fig-supp-gb-gatedep}.
These data show that there is no additional sign change for higher carrier densities indicating that the Seebeck coefficient at the grain boundary $\SGB$ is smaller in magnitude than the Seebeck coefficient of pristine graphene $\SG$ for the measurable range of carrier densities.
The measurements where done on the same device and grain boundary as the ones shown in the main text.
We note that  in the case of Fig.~\ref{fig-supp-gb-gatedep}c there is no sign change as a function of gate voltage visible as the charge neutrality point of the device was not reached due to high intrinsic doping.

When comparing the position of the photocurrent extrema in Fig.~\ref{fig-supp-gb-gatedep}c with the expected plasmonic fringe spacing as indicated by the dashed dotted curve it becomes evident indeed also the position of the photocurrent extrema is changing with carrier density. 
This could be due to an increased absorption due to the excitation of plasmons in the graphene and is subject of further study.

\subsection{Photodoping and puddles of encapsulated graphene}

\begin{figure*}[h]
\includegraphics{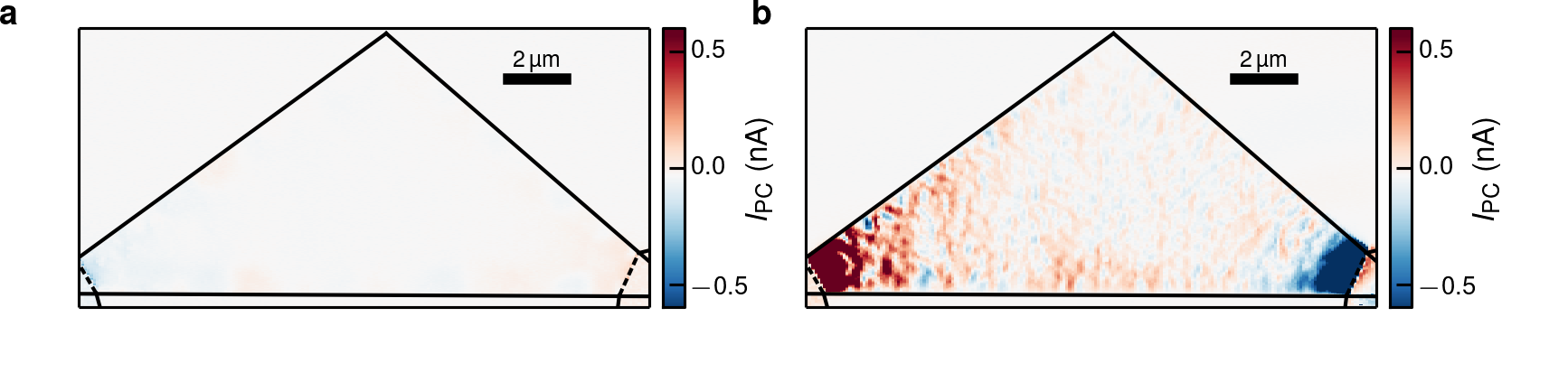}%
\caption{
\label{fig-supp-triangle}
\textbf{Encapsulated graphene device on oxidized silicon wafer.}
\textbf{a}, 
Near-field photocurrent of the graphene triangle device at -90~V backgate voltage in the dark.
\textbf{b},
Photodoping showing up in the near-field photocurrent after illuminating the device for several minutes with white LED light at $-90$~V. 
The dashed lines in \textbf{a} and \textbf{b} indicate where the stack is underneath the gold and the solid lines the outer edge of the stack and gold respectively.
}\end{figure*}

It is known that graphene on h-BN on top of an oxidized silicon wafer shows photodoping when it is illuminated with visible light.\cite{Ju2014}
This effect is clearly visible in our near-field photocurrent measurements after illumination with visible light. 
In Fig.~\ref{fig-supp-triangle}a we see almost no near-field photocurrent at a backgate voltage of $-90$~V.
In Fig.~\ref{fig-supp-triangle}b we show the near-field photocurrent pattern after the sample was illuminated with a white LED light source for several minutes at a backgate voltage of $-90$~V.
Clearly, near-field photocurrent is visible all the way throughout the device.
We attribute this to the screening of the backgate by photoexcited defects in the h-BN or at the h-BN/SiO2 interface, which can effectively neutralize the graphene.\cite{Ju2014}
Thus we see charge puddles which show a very similar near-field photocurrent pattern compared to the near-field photocurrent from charge puddles observed at the charge neutrality point of exfoliated graphene on SiO$_2$ shown in Fig.~3 and Fig.~4 of the main text.
In the case of the encapsulated device the charge puddles are induced by the photoexcited charged defects in the h-BN. 
In order to reset the photo doping and be immune to it we use positive gate voltages.\cite{Ju2014}

\subsection{Encapsulated graphene with a local gate}

\begin{figure*}[h]
\includegraphics{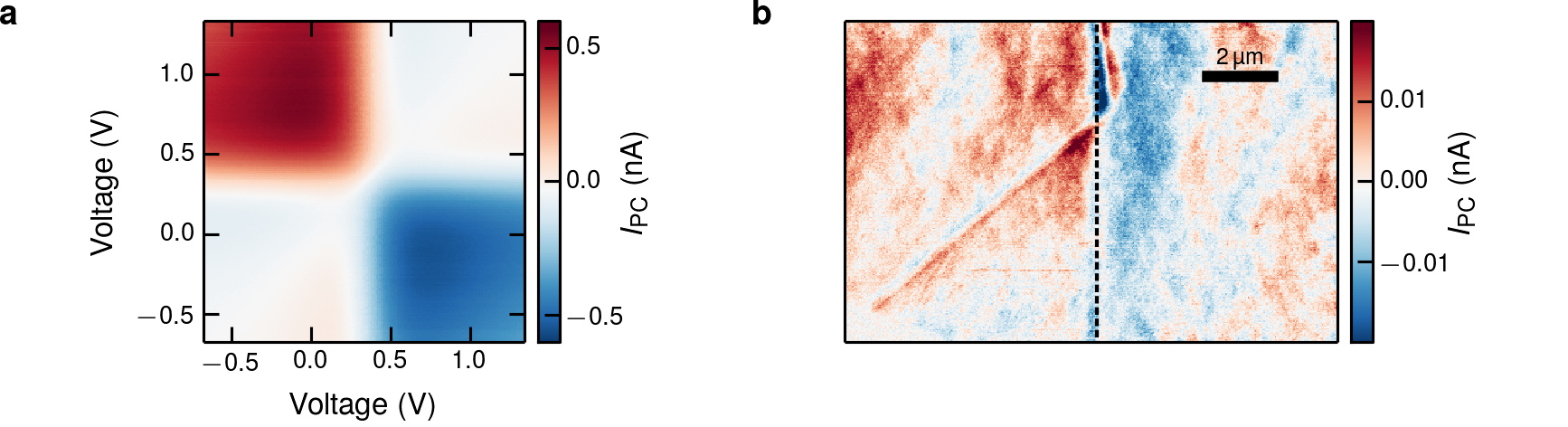}%
\caption{
\label{fig-supp-local-gate}
\textbf{Encapsulated graphene with independently tunable gates.}
\textbf{a}, 
6-fold pattern showing the thermoelectric origin of the near-field photocurrent at the junction between the two gates.\cite{Gabor2011}
\textbf{b},
Near-field photocurrent from a local gated encapsulated graphene device with both sides of the local gate being tuned to a voltage of $\Vbg = 0.4$~V close to the charge neutrality point.
The layers are AuPd(15~nm)/h-BN(42~nm)/Gr/h-BN(13~nm).
The dashed line indicates the position of the 50~nm gap between the two local gates.
}\end{figure*}

We also studied near-field photocurrent from an encapsulated device which was put onto conductive PdAu alloy gates with a 50~nm gap in between them in order to individually tune the carrier density in the graphene above the two gates.
In Fig.~\ref{fig-supp-local-gate}a we show a six-fold pattern typical for a photothermoelectric effect\cite{Gabor2011} in graphene measured at the junction between the two gates, which is indicated by the dashed line in Fig.~\ref{fig-supp-local-gate}.

In Fig.~\ref{fig-supp-local-gate}b the current in the device is shown for both sides being tuned to a gate voltage of 0.4~V, close to the charge neutrality point.
The magnitude of the near-field photocurrent is two orders of magnitude smaller than the magnitude of near-field photocurrent from charge puddles in graphene on SiO$_2$.
This indicates that the charge density inhomogeneity of the charge puddles in encapsulated graphene on a local gate is much lower than for graphene on SiO$_2$.
Furthermore the length scale of the charge puddles indicates that the charge puddle size is much larger than for the case of graphene on SiO$_2$\cite{Xue2011,Decker2011} or that the cooling length in encapsulated graphene devices is much longer than for bare graphene, as expected due to the increased carrier mobility in encapsulated graphene.

We remark the we do not see any photodoping for the encapsulated graphene on top of a conductive gate structure, even after extended periods of exposure of many minutes to the same white light LED as was used to induce photodoping in Fig.~\ref{fig-supp-triangle}b.
This can be explained by the extraction of the photoexcited charged defects by the conductive gate, which is in direct contact with h-BN.

\section{Modelling}
In this section we give a more detailed overview of the photothermoelectric photocurrent model used to describe the measurements and models in the main text.

\subsection{Photothermoelectric model}
Near-field photocurrent $\Ipc$ in graphene is governed by the photothermoelectric effect, which can be calculated by:\cite{Song2011b,Gabor2011,Tielrooij2014}

\begin{equation}
\label{eq-PTE}
\Ipc(x,y) = \frac{1}{RW} \int \dfrac{\partial T(x,y)}{\partial x} S(x,y) \, dxdy
\end{equation}

where $R$ is the total resistance, consisting of graphene, contact and circuit resistance, $W$ the width of the device, $\partial T/\partial x$ is the gradient of the temperature $T$ in current flow direction $x$ and $S$ is the Seebeck coefficient.
In order to simulate the near-field photocurrent in arbitrary geometries special care has to be taken.\cite{Song2014} 

For calculating $\Ipc$ at each position one has to do a convolution of the spatial Seebeck coefficient profile with the temperature gradient.
This is numerically very expensive and the simulations can take a long time.
In order to more efficiently simulate a near-field photocurrent map for an arbitrary spatial Seebeck profile we use the convolution theorem.
This allows us to just multiply the Fourier transform of the spatial Seebeck coefficient and temperature gradient, multiply them and inverse Fourier transform them.
This is computationally much faster and the simulation time is greatly reduced.
This method was employed for all simulations shown in the main text as well as the Supplement.

\subsection{Heat spreading in 2D}
In order to describe the near-field photocurrent in graphene with a model it is of great importance to correctly describe the heat profile within the graphene as the heat profile in combination with the Seebeck profile lead to the near-field photocurrent pattern.

Graphene on top of a substrate is a two dimensional material with the substrate acting as a heat sink, as for typical substrate materials such as SiO$_2$ the thermal conductivity is much larger than for air. 
The two dimensional heat equation in steady state with an additional term for heat sinking can be written as:

\begin{equation}
\label{eq-2d-heat-equation}
0 = \kappa \nabla^2 T + P - g (T - T_\mathrm{s}),
\end{equation}

where $T$ is the electron temperature, $T_\mathrm{s}$ is the constant heat sink temperature, $\kappa$ the thermal conductivity in plane, $g$ the thermal conductivity out of plane to the heat sinking substrate and $P$ the power density of the heat source.

\begin{figure*}[h]
\includegraphics{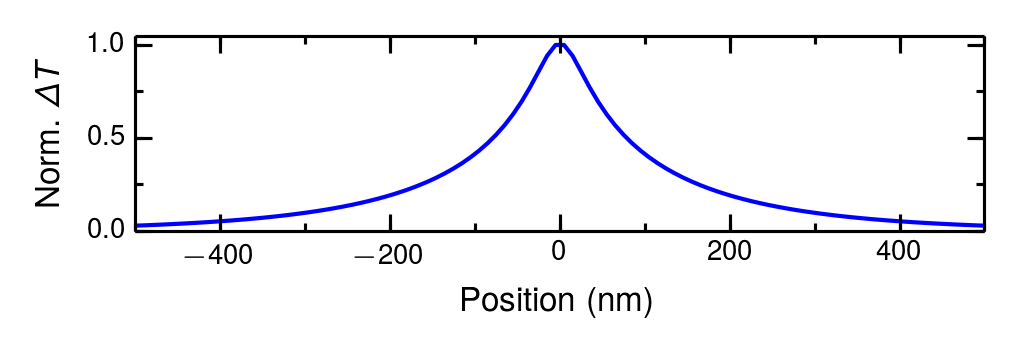}%
\caption{
\label{fig-supp-tempspot}
\textbf{Steady state temperature profile}
created in infinite graphene with a thermal length of 200~nm and a finite tip size correction of 25~nm.
}\end{figure*}

We apply this equation by assuming that the electron heat cools to
the lattice, but then that the lattice can pass on any heat much more
easily to substrate. 
In that way the temperature of the lattice does not change significantly. 
This is convenient because the lattice itself has a different thermal length which would complicate matters.

If we take the two-dimensional case and a point source for $P$,

\begin{equation}
P = P_\mathrm{total} \delta(x)\delta(y),
\end{equation}
then the solution of the two dimensional heat eq.~\ref{eq-2d-heat-equation} turns out to be:
\begin{equation}
T_\mathrm{spot} = T_0  K_0\left(\dfrac{|r|}{l_\mathrm{cool}}\right)
\end{equation}

where $K_0$ is the modified Bessel function of the second kind, $r = \sqrt{x^2+y^2}$, the cooling length $l_\mathrm{cool} = \sqrt{\kappa/g}$ and the maximum temperature rise $T_0 = P_\mathrm{total} / (2\pi\kappa)$.

If we also include $l_\mathrm{tip}$ to approximate the effect of the finite tip size we end up with a heat spot of the following form:

\begin{equation}
\label{eq-heatspot}
T_\mathrm{spot}\left(x'-x,y'-y\right) = T_0 K_0\left(\sqrt{\dfrac{(x'-x)^2+(y'-y)^2+l_\mathrm{tip}^2}{l_\mathrm{cool}^2}}\right)
\end{equation}

A typical temperature profile as calculated by eq.~\ref{eq-heatspot} with a finite tip size approximation $l_\mathrm{tip} = 25$~nm, the actual radius of the tip, and a cooling length of $l_\mathrm{cool} = 200$~nm is presented in Fig.~\ref{fig-supp-tempspot}.

\subsection{Grain boundary model}
\label{gb-model}
\begin{figure*}[h]
\includegraphics{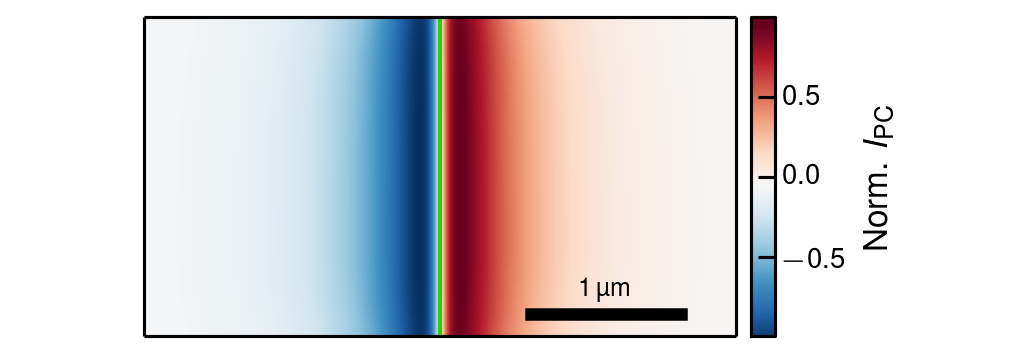}%
\caption{
\label{fig-supp-gb-map}
\textbf{Simulated near-field photocurrent map from a grain boundary.}
A map of the simulated near-field photocurrent for each position of the graphene device.
The contacts are on the left and right outside of the region shown in this figure.
The cooling length used was 140~nm and the tip size 25~nm.
}\end{figure*}

Grain boundaries can be modelled as having a finite width with a Gaussian profile\cite{Fei2013} and their Seebeck coefficient is smaller in magnitude than the one of the surrounding pristine graphene, in accordance with the results of the main text.
The near-field photocurrent profile is then calculated by performing a two dimensional convolution between the temperature profile defined by eq.~\eqref{eq-heatspot} and the Seebeck profile.
The results of this convolution are shown in Fig.~\ref{fig-supp-gb-map}.

\subsection{Seebeck coefficient of polycrystalline graphene}
\label{seebeck-Roche}

We use an order-N Kubo-Greenwood wavepacket approach to calculate the conductivity of polycrystalline samples with different average grain sizes,\cite{VanTuan2013} and we use square samples to convert this conductivity to conductance ($G$). 
Additionally, we calculate the conductance of pristine graphene using a Landauer approach. 
The Seebeck coefficient is calculated as the ratio of the first- and zero-order Onsager coefficients,

\begin{align}
S(\mu,T) = -\frac{1}{|e|T} \frac{\int_{-\infty}^{\infty} (E-\mu)G(E)\bigg(-\frac{\partial f}{\partial E}\bigg) dE }{\int_{-\infty}^{\infty} G(E)\bigg(-\frac{\partial f}{\partial E}\bigg) dE },
\end{align}
where $f$ is the Fermi-Dirac distribution, $\mu$ the chemical potential and we set the temperature $T$ to 300 K.

\begin{figure}[h]
\centering
\includegraphics{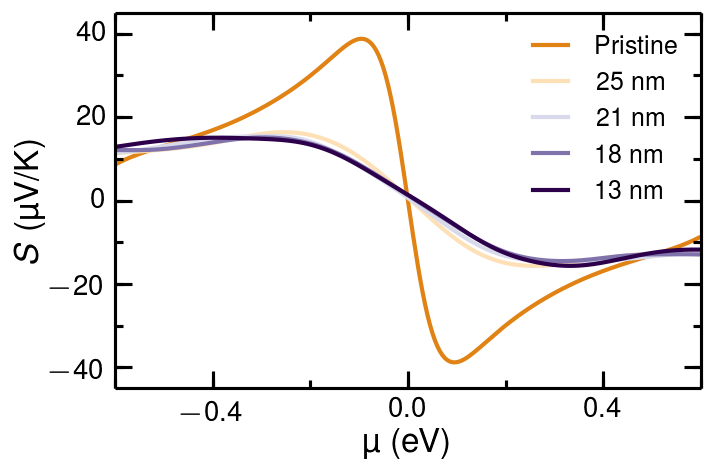}
\caption{Simulated Seebeck coefficient for different polycrystalline samples ($T = 300$~K), for grain sizes (13, 18, 21 and 25 nm).}
\label{Seebeck}
\end{figure}

For polycrystalline samples with different average grain size (13, 18, 21 and 25~nm) we find that the Seebeck coefficient is significantly reduced compared to the clean case due to scattering at the grain boundaries (Figure~\ref{Seebeck}). 
Furthermore, the Seebeck coefficient is independent of grain size, resulting from the linear scaling of charge transport in polycrystalline graphene \cite{VanTuan2013,Cummings2014}.
The impact of the grain boundaries is reduced for larger chemical potentials ($>0.3$~eV), where the Seebeck coefficient for polycrystalline and pristine graphene is similar.

The simulation of the backgate dependent near-field photocurrent in Fig.~2e of the main text was performed by using a Gaussian by using a Gaussian spatial distribution of the Seebeck coefficient with a full width at half maximum of 20~nm.
The difference in Seebeck coefficient between $\SG$ of the pristine graphene and $\SGB$ at the grain boundary was calculated for each measured backgate voltage.
For $\SGB$, the simulation corresponding to 25~nm polycrystalline graphene was used, but the results are essentially the same for different grain sizes.
The spatial near-field photocurrent profile for each backgate voltage was calculated according to the procedure given in \ref{gb-model} and then normalized by the resistance $R$ for each gate voltage according to eq.~\eqref{eq-PTE}.

\subsection{Charge puddle model}
In order to model near-field photocurrent from charge puddles, we first need to find an accurate model of the charge puddle distribution.
For this we use a random spatial Seebeck profile generated by a spatial profile of white noise and smoothing the noise to create charge puddles with an average approximate size of 20~nm.
This size was extracted from previous experimental studies of charge puddles on SiO$_2$.\cite{Chen2008,Zhang2009a,Decker2011,Xue2011} 

\begin{figure*}[h]
\includegraphics{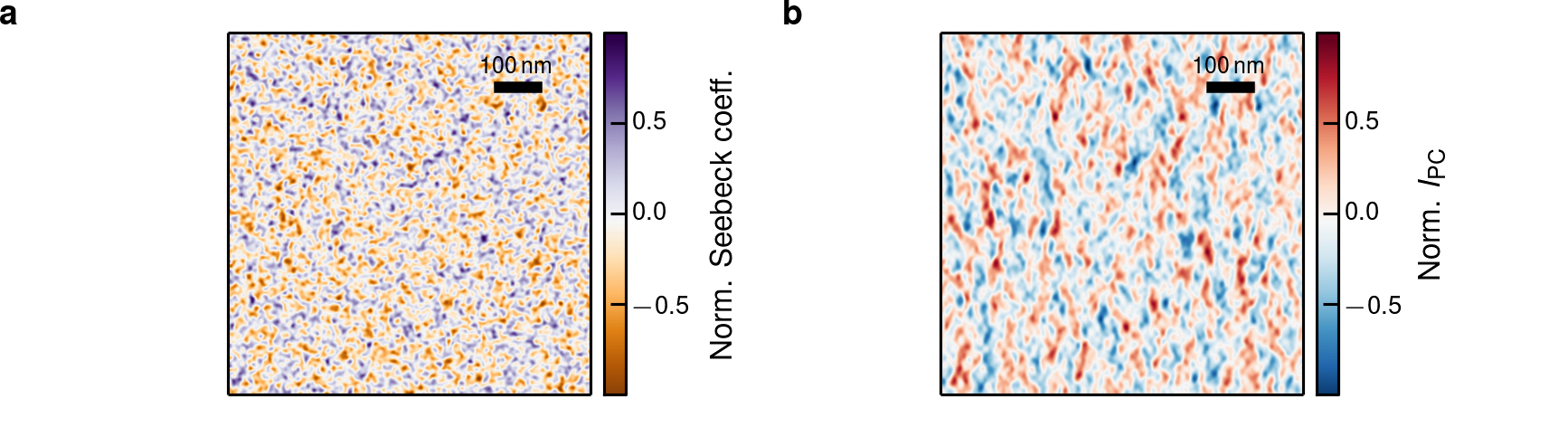}%
\caption{
\label{fig-supp-puddles-theory}
\textbf{Simulation of a near-field phtocurrent map induced by a random charge puddle distribution.}
\textbf{a}, 
Random distribution of Seebeck coefficients due to a random distribution of charge puddles with an average size of $\sim$~20~nm.
\textbf{b},
Near-field photocurrent generated by the charge puddles in \textbf{a} with a cooling length of the charge carriers in the graphene of 200~nm and a tip size of 25~nm.
}\end{figure*}

We calculate the near-field photocurrent map from the charge puddles.
To this end, we again convolve the spatial Seebeck profile with the spatial heat gradient profile.
A comparison between the spatial Seebeck profile and the near-field photocurrent map is shown in Fig.~\ref{fig-supp-puddles-theory}.
Here the source and drain contacts are outside of the display on the left and right respectively.
It is obvious that for positions with a high gradient in Seebeck coefficient the near-field photocurrent is strongest, as expected from eq.~\ref{eq-PTE}.

\subsection{Model of a \textit{pnp}-junction}
\begin{figure*}[h]
\includegraphics{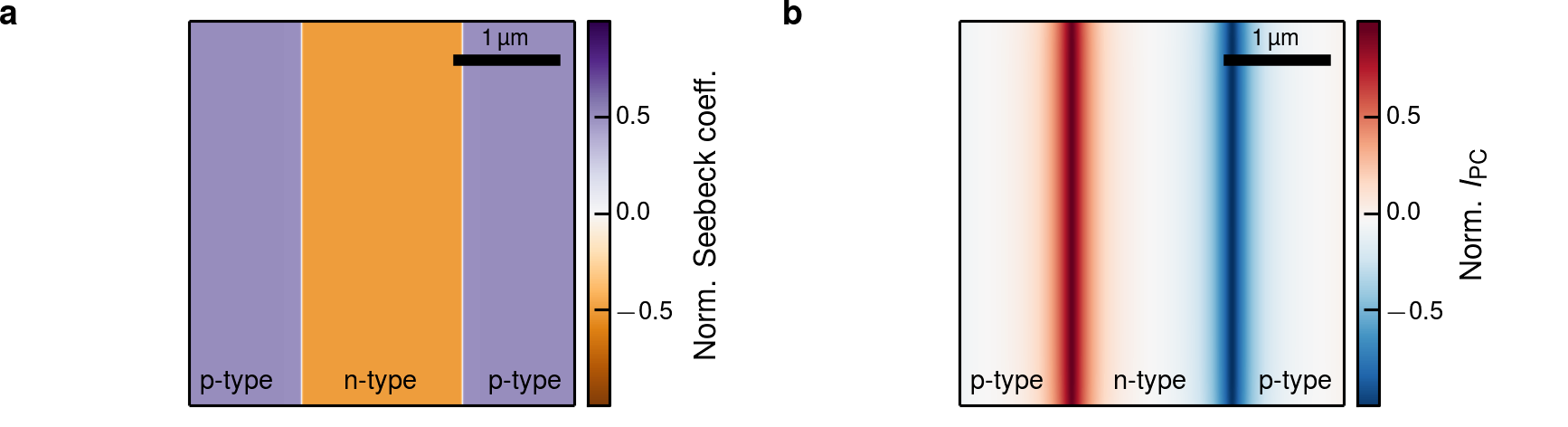}%
\caption{
\label{fig-supp-pnp-junction}
\textbf{Simulation of a near-field photocurrent map from a pnp-junction.}
\textbf{a}, 
Spatial Seebeck coefficient profile for two p- and one n-type regions.
The doping in the two regimes is assumed to be of the same magnitude and opposite sign.
\textbf{b},
Near-field photocurrent generated by the $pn$- and $np$-junction with a cooling length of 200~nm and a tip size of 25~nm.
}\end{figure*}

Here, we model the near-field photocurrent map of a $pnp$-junction, where the $p$-type and $n$-type regions are extended to a size larger than the heat spot size.
The transition length scale of the regions is smaller than the heat spot size.
In this case the near-field photocurrent at the $pn$- and $np$-junction respectively has a single sign as observed for example in the triangular near-field photocurrent pattern due to edge doping in Fig.~5f in the main text.
In the simulations shown in Fig.~\ref{fig-supp-pnp-junction}a and b the electrical contacts are on the left and right outside of the shown region.

\end{document}